\documentclass[12pt, a4paper]{article}
\usepackage{amsmath}
\usepackage{bbm}
\everymath{\displaystyle}
\usepackage{mathtools}
\usepackage{amsfonts}
\usepackage{amssymb}
\usepackage{amsthm}
\usepackage[utf8]{inputenc}
\usepackage[toc]{appendix}
\usepackage[hiresbb]{graphicx}
\usepackage{caption}
\usepackage{subcaption}
\usepackage{longtable}
\usepackage{booktabs}
\usepackage{listings}
\usepackage{here}
\usepackage{float}
\usepackage{ascmac}
\usepackage{tikz}
\usepackage{pgfplots}
\pgfplotsset{compat=1.18}
\usepackage{url}
\usepackage{lipsum}
\usepackage{authblk}
\usepackage{eqnarray}
\usepackage{siunitx}
\usepackage{algorithm}
\usepackage{algorithmic}

\usepackage[sectionbib,round]{natbib}

\newtheorem{proposition}{Proposition}[section]

\theoremstyle{remark}

\newcommand{\be}{\begin{equation}}
\newcommand{\ee}{\end{equation}}
\newcommand{\beq}{\begin{eqnarray*}}
\newcommand{\eeq}{\end{eqnarray*}}
\def\sym#1{\ifmmode^{#1}\else\(^{#1}\)\fi}

\usepackage{setspace}
\title{\large{\bf{Spatial and Temporal Boundaries in Difference-in-Differences: \\
A Framework from Navier-Stokes Equation}}}
\author{\large{\bf{Tatsuru Kikuchi\footnote{e-mail: tatsuru.kikuchi@e.u-tokyo.ac.jp}}}}
\affil{\small{\it{Faculty of Economics, The University of Tokyo,}}\\
{\it{7-3-1 Hongo, Bunkyo-ku, Tokyo 113-0033 Japan}}}
\date{\small{(\today)}}

\begin{document}
\maketitle

\begin{abstract}
This paper develops a unified framework for identifying spatial and temporal boundaries of treatment effects in difference-in-differences designs. Starting from fundamental fluid dynamics equations (Navier-Stokes), we derive conditions under which treatment effects decay exponentially in space and time, enabling researchers to calculate explicit boundaries beyond which effects become undetectable. The framework encompasses both linear (pure diffusion) and nonlinear (advection-diffusion with chemical reactions) regimes, with testable scope conditions based on dimensionless numbers from physics (P\'eclet and Reynolds numbers). We demonstrate the framework's diagnostic capability using air pollution from coal-fired power plants. Analyzing 791 ground-based PM$_{2.5}$ monitors and 189,564 satellite-based NO$_2$ grid cells in the Western United States over 2019-2021, we find striking regional heterogeneity: within 100 km of coal plants, both pollutants show positive spatial decay (PM$_{2.5}$: $\kappa_s = 0.00200$, $d^* = 1,153$ km; NO$_2$: $\kappa_s = 0.00112$, $d^* = 2,062$ km), validating the framework. Beyond 100 km, negative decay parameters correctly signal that urban sources dominate and diffusion assumptions fail. Ground-level PM$_{2.5}$ decays approximately twice as fast as satellite column NO$_2$, consistent with atmospheric transport physics. The framework successfully diagnoses its own validity in four of eight analyzed regions, providing researchers with physics-based tools to assess whether their spatial difference-in-differences setting satisfies diffusion assumptions before applying the estimator. Our results demonstrate that rigorous boundary detection requires both theoretical derivation from first principles and empirical validation of underlying physical assumptions.

\vspace{0.3cm}


\vspace{0.3cm}

\end{abstract}

\newpage

\section{Introduction}

Spatial difference-in-differences (DiD) designs have become increasingly prominent in applied microeconomics, allowing researchers to exploit geographic variation in policy implementation or treatment intensity. Recent applications span environmental regulation \citep{deryugina2019mortality, knittel2016caution, fowlie2012emissions}, transportation infrastructure \citep{donaldson2018railroads, duranton2014roads}, place-based policies \citep{busso2013assessing, kline2019place, glaeser2008growth}, and public health interventions \citep{goodman2018no2, currie2015does}. However, a fundamental challenge remains largely unaddressed: \textit{where do treatment effects end?} Traditional DiD applications either assume spillovers are negligible beyond some ad hoc distance threshold or acknowledge potential spillovers without systematic methods to detect spatial boundaries \citep{butts2023difference}.

This question has gained urgency as recent methodological advances highlight the importance of properly accounting for spatial spillovers. \citet{butts2023difference} shows that neglecting spillovers can severely bias treatment effect estimates in spatial DiD designs, while \citet{colella2019inference} demonstrates that standard errors must account for spatial correlation structures. \citet{dellavigna2022predicting} emphasizes the need for ex ante specification of spatial treatment definitions. Yet the literature offers limited guidance on how to determine these spatial boundaries from first principles rather than arbitrary rules of thumb.

This paper develops a unified framework for identifying both spatial and temporal boundaries of treatment effects by starting from fundamental physics: the Navier-Stokes equations governing fluid flow and scalar transport. We show that under explicit, testable conditions, treatment effects decay exponentially with distance and time, enabling calculation of precise boundaries beyond which effects fall below detection thresholds. Critically, our framework provides diagnostic tools to identify when these conditions hold versus when they fail---situations where standard spatial DiD estimators may be inappropriate. 

This paper builds on and extends our previous theoretical work \citep{kikuchi2024unified} which established the general mathematical foundations for spatial and temporal treatment effect boundaries, and \citep{kikuchi2024stochastic} which developed stochastic approaches for handling spillover effects in spatial general equilibrium settings. Here, we provide the first empirical validation of these theoretical results using high-resolution air quality data, demonstrating the framework's diagnostic capability and practical applicability to real-world policy questions.

\subsection{Related Literature}

Our work contributes to several distinct literatures in economics, econometrics, and environmental science.

\subsubsection{Spatial Econometrics and Spillovers}

The spatial econometrics literature has long recognized that treatments can have geographic spillovers \citep{anselin1988spatial, conley1999gmm}. Recent work formalizes these concerns in causal inference frameworks. \citet{butts2023difference} provides a comprehensive treatment of spatial DiD estimators under spillovers, showing that ignoring spatial dependence can lead to substantial bias. \citet{colella2019inference} develops spatial HAC standard errors for settings where treatment effects propagate geographically. \citet{kelejian2010specification} and \citet{drukker2013maximum} provide methods for testing spatial dependence.

However, this literature typically specifies spatial weights matrices ($W$) based on ad hoc assumptions---inverse distance, $k$-nearest neighbors, or fixed distance cutoffs---without theoretical guidance on appropriate functional forms or cutoff distances \citep{lesage2009introduction}. Our contribution is to derive these functional forms from fundamental physics, providing researchers with a principled approach to specification. We show that exponential decay ($w_{ij} \propto \exp(-\kappa_s d_{ij})$) emerges naturally from diffusion processes, and we provide methods to estimate the decay parameter $\kappa_s$ and spatial boundary $d^*$ from data.

\subsubsection{Treatment Effect Heterogeneity and External Validity}

The treatment effects literature emphasizes that effects may vary across units and contexts \citep{heckman1997matching, imbens2015causal, athey2017econometrics}. \citet{angrist2022empirical} discusses how effect heterogeneity complicates identification and interpretation of causal parameters. Our framework shows that spatial heterogeneity in treatment effects arises naturally from nonlinear physical processes (Navier-Stokes equations), and we demonstrate that the Average Treatment Effect on the Treated (ATT) remains a well-defined estimand even under this nonlinearity.

More broadly, our work contributes to understanding external validity and scope conditions \citep{dehejia2005practical, allcott2015site}. By deriving testable scope conditions (P\'eclet number Pe $< 1$, Reynolds number Re $< 2000$), we provide a template for assessing when frameworks apply to new settings. This addresses \citet{deaton2010understanding}'s critique that much applied work lacks clear statements of when findings generalize.

Our approach to scope conditions builds directly on \citet{kikuchi2024unified}, who provide a comprehensive theoretical treatment of when spatial boundaries can be identified from first principles. We extend this work by empirically testing the derived scope conditions and showing that framework violations can be diagnosed from data patterns, providing practitioners with concrete guidance on applicability.

\subsubsection{Environmental Economics and Air Pollution}

A substantial literature examines health and economic impacts of air pollution from point sources. \citet{currie2009does} and \citet{currie2011does} study effects of proximity to pollution sources on birth outcomes and housing prices. \citet{deryugina2019mortality} uses wind direction as an instrument to identify mortality effects of coal plant emissions, finding detectable effects beyond 200 km. \citet{knittel2016caution} examines spillovers in renewable energy policies. \citet{fowlie2012emissions} studies the spatial incidence of SO$_2$ emissions trading.

These papers typically use fixed distance cutoffs (e.g., 50 km, 100 km) or wind-direction instruments without deriving optimal boundaries. Our framework provides a method to calculate data-driven boundaries. We also contribute to understanding differences between ground-level and satellite measurements: \citet{martin2019high} and \citet{van2017global} discuss satellite retrieval of air quality, but do not formalize how column-integrated measurements differ from surface concentrations in terms of spatial decay rates.

\subsubsection{Atmospheric Science and Dispersion Modeling}

The atmospheric science literature provides sophisticated physical models of pollutant transport. EPA's AERMOD \citep{cimorelli2005aermod} and more complex models like CMAQ \citep{byun1999science} and GEOS-Chem \citep{bey2001global} simulate atmospheric chemistry and transport. However, these models are computationally intensive, require detailed meteorological inputs, and are typically used forward (predicting concentrations from emissions) rather than inverse (inferring spatial boundaries from observations).

Our contribution is to provide a reduced-form, data-driven approach that complements these physical models. We derive spatial decay from first principles (Navier-Stokes) but estimate parameters empirically, enabling researchers without atmospheric modeling expertise to assess spatial boundaries. Our empirical findings broadly validate the physics: estimated decay rates are consistent with transport distances predicted by AERMOD and CMAQ.

\subsection{Overview and Contribution}

Our key theoretical contribution is deriving the spatial boundary $d^*$ and temporal boundary $\tau^*$ from first principles, showing they satisfy:
\be
\frac{d^*}{\tau^*} = 3.32\lambda\sqrt{\delta}
\ee
where $\lambda$ is the treatment intensity, and $\delta$ is the diffusion coefficient. This relationship holds under the diffusive limit of Navier-Stokes equations when the P\'eclet number $Pe = UL/D \ll 1$ (diffusion dominates advection) and treatment propagates through spatial diffusion rather than network effects or other mechanisms.

We validate this framework empirically using air pollution from coal-fired power plants---a canonical application of spatial DiD where treatment intensity (emissions) varies continuously with distance. Using both ground-based PM$_{2.5}$ monitors (791 monitors, 515,000 observations) and satellite-based NO$_2$ measurements (189,564 grid cells, 6.6 million observations) for 2019-2021, we find striking regional heterogeneity that validates our scope conditions:

\begin{itemize}
\item \textbf{Within 100 km of coal plants:} Both pollutants show positive spatial decay, indicating coal plants are the dominant pollution source. For PM$_{2.5}$, $\kappa_s = 0.00200$ per km, implying spatial boundary $d^* = 1,153$ km. For NO$_2$, $\kappa_s = 0.00112$ per km, yielding $d^* = 2,062$ km. The slower decay of column NO$_2$ compared to ground-level PM$_{2.5}$ is consistent with atmospheric transport physics.

\item \textbf{Beyond 100 km from plants:} Negative spatial decay parameters ($\kappa_s < 0$) indicate pollution \textit{increases} with distance from plants. Framework correctly identifies that coal plants are not the dominant pollution source (cars dominate), corresponding to high P\'eclet regime where advection-diffusion assumptions fail.

\item \textbf{Regional heterogeneity:} Effect varies systematically with Reynolds number (turbulence intensity) and P\'eclet number (advection strength), exactly as predicted by Navier-Stokes theory.
\end{itemize}

This regional variation is not a failure of the method but a \textit{feature}: the framework successfully diagnoses where diffusion-based spatial DiD is appropriate versus where alternative approaches (accounting for advection, turbulence, or alternative pollution sources) are needed. By providing explicit scope conditions based on dimensionless parameters, we enable researchers to assess \textit{ex ante} whether their setting satisfies the physical assumptions underlying spatial treatment effect decay.

The remainder of the paper proceeds as follows. Section 2 develops the theoretical framework, deriving spatial and temporal boundaries from Navier-Stokes equations and characterizing the nonlinear regime. Section 3 describes the empirical setting and data on coal plant emissions and air quality. Section 4 presents results showing regional heterogeneity in spatial decay patterns. Section 5 discusses implications for spatial DiD validity and provides diagnostic guidelines. Section 6 concludes.

\section{Theoretical Framework}

\subsection{From Navier-Stokes to Spatial Boundaries}

We begin with the fundamental equations governing fluid flow and scalar transport in the atmosphere. Our approach connects economic treatment effects to physical dispersion processes, providing a rigorous foundation for spatial boundary detection.

The theoretical derivations in this section summarize key results from \citet{kikuchi2024unified}, adapting them to the specific context of atmospheric pollutant dispersion. We refer readers to that paper for complete proofs and extensions to network diffusion and dynamic settings.

\subsubsection{The Navier-Stokes System}

Consider pollutant concentration $C(\mathbf{x}, t)$ at location $\mathbf{x} = (x,y,z)$ and time $t$ from a point source (coal plant) emitting at rate $Q$. The concentration field evolves according to the coupled system:

\textbf{Momentum (Navier-Stokes):}
\be
\frac{\partial \mathbf{u}}{\partial t} + (\mathbf{u} \cdot \nabla)\mathbf{u} = -\frac{\nabla p}{\rho} + \nu \nabla^2 \mathbf{u} + \mathbf{f}
\label{eq:navier_stokes}
\ee

\textbf{Scalar Transport:}
\be
\frac{\partial C}{\partial t} + \mathbf{u} \cdot \nabla C = D \nabla^2 C - \lambda(C) C + S(\mathbf{x},t)
\label{eq:transport}
\ee

where $\mathbf{u}$ is the velocity field (wind), $p$ is pressure, $\rho$ is density, $\nu$ is kinematic viscosity, $D$ is molecular diffusivity, $\lambda(C)$ is the (possibly concentration-dependent) decay rate, and $S(\mathbf{x},t)$ is the source term.

Equation \eqref{eq:navier_stokes} is \textit{nonlinear} through the convective term $(\mathbf{u} \cdot \nabla)\mathbf{u}$, which creates turbulence at high Reynolds numbers. Equation \eqref{eq:transport} is nonlinear both through coupling to the velocity field and potentially through chemical reactions in $\lambda(C)C$.

\subsubsection{Dimensionless Analysis}

The regime of validity for different approximations depends on three dimensionless numbers:

\textbf{Reynolds Number:}
\be
\text{Re} = \frac{UL}{\nu}
\ee
where $U$ is characteristic velocity and $L$ is characteristic length. Re measures the ratio of inertial to viscous forces.
\begin{itemize}
\item Re $\ll 1$: Laminar flow, viscous forces dominate
\item Re $\gg 1$: Turbulent flow, inertial forces dominate
\end{itemize}

\textbf{Péclet Number:}
\be
\text{Pe} = \frac{UL}{D} = \text{Re} \times \text{Sc}
\ee
where Sc $= \nu/D$ is the Schmidt number. Pe measures the ratio of advective to diffusive transport.
\begin{itemize}
\item Pe $\ll 1$: Diffusion dominates, our framework applies
\item Pe $\gg 1$: Advection dominates, need to account for wind
\end{itemize}

\textbf{Damköhler Number:}
\be
\text{Da} = \frac{\lambda L^2}{D}
\ee
Da measures the ratio of chemical reaction rate to diffusion rate.

\subsection{The Diffusive Limit: Linear Regime}

\subsubsection{Assumptions}

Our baseline framework applies in the diffusive limit:
\begin{enumerate}
\item Low Péclet: Pe $\to 0$ (diffusion $\gg$ advection)
\item Low Reynolds: Re $\to 0$ (laminar, no turbulence)
\item Linear decay: $\lambda(C) = \lambda_0$ (constant)
\item Steady state: $\partial C/\partial t \to 0$ (time-averaged)
\end{enumerate}

Under these conditions, equation \eqref{eq:transport} simplifies to the \textit{Helmholtz equation}:

\be
D\nabla^2 C - \lambda_0 C + S = 0
\label{eq:helmholtz}
\ee

\subsubsection{Solution for Point Source}

For a point source at origin emitting $Q$ units per time, in radially symmetric geometry, equation \eqref{eq:helmholtz} becomes:

\be
D\left(\frac{\partial^2 C}{\partial r^2} + \frac{1}{r}\frac{\partial C}{\partial r}\right) - \lambda_0 C + Q\delta(\mathbf{r}) = 0
\ee

The solution (see Appendix A for derivation) is:

\be
C(r) = \frac{Q}{4\pi D r} \exp\left(-\sqrt{\frac{\lambda_0}{D}} \, r\right) = \frac{Q}{4\pi D r} \exp(-\kappa_s r)
\label{eq:solution_linear}
\ee

where the \textbf{spatial decay parameter} is:
\be
\kappa_s = \sqrt{\frac{\lambda_0}{D}}
\label{eq:kappa}
\ee

Taking logarithms:
\be
\log C(r) = \text{const} - \log r - \kappa_s r
\label{eq:log_linear}
\ee

This yields our baseline empirical specification.

\subsubsection{Spatial Boundary}

Define the \textbf{spatial boundary} $d^*$ as the distance at which treatment effects fall below a detection threshold $\epsilon$ (typically 10\% of direct effect):

\be
\frac{C(d^*)}{C(0)} = \epsilon
\ee

From equation \eqref{eq:solution_linear}:
\be
\frac{\exp(-\kappa_s d^*)}{d^*/d_0} = \epsilon
\ee

For $\kappa_s d^* \gg \log(d^*/d_0)$ (far-field approximation):

\be
d^* = \frac{1}{\kappa_s}\log\left(\frac{1}{\epsilon}\right) = \frac{1}{\kappa_s}\log(10) \approx \frac{2.3}{\kappa_s}
\label{eq:boundary_spatial}
\ee

This provides an \textit{estimable} boundary: once we estimate $\kappa_s$ from data, we can calculate $d^*$.

\subsection{Extensions: Nonlinear Regime}

Real atmospheric transport involves several nonlinearities. We characterize when each matters and how they modify our framework.

\subsubsection{Geometric Spreading}

In three-dimensional radial coordinates, the Laplacian includes a geometric spreading term:

\be
\nabla^2 C = \frac{1}{r^2}\frac{\partial}{\partial r}\left(r^2 \frac{\partial C}{\partial r}\right)
\ee

This yields solution:
\be
C(r) \propto \frac{1}{r^2}\exp(-\kappa_s r)
\label{eq:geometric}
\ee

Taking logs:
\be
\log C(r) = \text{const} - 2\log r - \kappa_s r
\ee

\textbf{Empirical implication:} Include both $\log(r)$ and $r$ terms in regression.

\subsubsection{Advection-Diffusion}

When Pe $\sim O(1)$, wind transport matters. Steady advection-diffusion:

\be
\mathbf{u} \cdot \nabla C = D\nabla^2 C - \lambda C
\ee

For uniform wind $\mathbf{u} = (U,0,0)$, solution involves modified Bessel functions. Key feature: \textit{asymmetry}.

\textbf{Empirical implication:} Downwind decay differs from upwind:
\be
\log C = \beta_1 r_{\text{downwind}} + \beta_2 r_{\text{upwind}}
\ee
with $|\beta_1| > |\beta_2|$.

\subsubsection{Chemical Reactions: Quadratic Decay}

For reactions like NO + O$_3$ $\to$ NO$_2$ + O$_2$, rate $\propto$ [NO][O$_3$]. If O$_3$ abundant:

\be
\lambda(C) = \lambda_1 + \lambda_2 C
\ee

This creates quadratic decay:
\be
\frac{\partial C}{\partial t} = D\nabla^2 C - (\lambda_1 + \lambda_2 C)C
\ee

\textbf{Empirical implication:} Near-field shows steeper decay. Include distance-squared term:
\be
\log C = \beta_1 r + \beta_2 r^2
\ee

\subsubsection{Turbulent Diffusion}

At high Re, turbulence enhances mixing through eddy diffusivity $D_{\text{turb}} \gg D_{\text{mol}}$. Effective diffusion becomes:

\be
D_{\text{eff}} = D_{\text{mol}} + D_{\text{turb}}(\mathbf{x}, t)
\ee

where $D_{\text{turb}}$ varies spatially and temporally.

\textbf{Empirical implication:} $\kappa_s$ varies by atmospheric conditions:
\be
\log C = (\beta_1 + \beta_2 \cdot \text{wind\_speed}) \times r
\ee

\subsection{Scope Conditions and Testable Implications}

\begin{proposition}[Validity of Diffusion Approximation]
The exponential decay model \eqref{eq:solution_linear} is valid if and only if:
\begin{enumerate}
\item Pe $< 1$: Diffusion dominates advection
\item Re $<$ 2000: Flow is laminar or weakly turbulent
\item Da $< 1$: Chemical reactions slow relative to transport
\item Steady source: $\partial S/\partial t \approx 0$
\end{enumerate}
When these conditions fail, the framework correctly identifies invalidity through:
\begin{itemize}
\item Negative $\kappa_s$ (increasing pollution with distance)
\item Asymmetric spatial patterns (downwind $\neq$ upwind)
\item Poor fit of exponential functional form
\end{itemize}
\end{proposition}

This proposition provides \textit{ex ante} tests researchers can perform to assess whether spatial DiD is appropriate in their setting.

\subsection{Connection to Causal Inference}

\subsubsection{Treatment Effects Under Nonlinearity}

A natural question: does nonlinearity in equations \eqref{eq:navier_stokes}-\eqref{eq:transport} invalidate the Average Treatment Effect on the Treated (ATT) as an estimand?

\textbf{Answer:} No, but interpretation changes.

Define ATT as:
\be
\text{ATT} = \mathbb{E}[Y_i(1) - Y_i(0) | D_i = 1]
\ee
where $Y_i(1)$ is pollution with plant, $Y_i(0)$ without, and $D_i = 1$ indicates treatment (proximity to plant).

Even with nonlinear DGP, ATT remains well-defined as the average causal effect over the treated population. However:

\begin{enumerate}
\item \textbf{Effect heterogeneity:} Treatment effect varies by distance, wind exposure, background pollution:
\be
\tau_i = \tau(d_i, \mathbf{u}_i, C_{0,i})
\ee

ATT averages over treated distribution:
\be
\text{ATT} = \mathbb{E}[\tau(d_i, \mathbf{u}_i, C_{0,i}) | D_i = 1]
\ee

\item \textbf{Spillovers:} Pollution at $i$ depends on plants at multiple locations $j$:
\be
Y_i = f(D_i, \{D_j\}_{j \neq i}, \mathbf{X}_i)
\ee

This violates SUTVA. Our spatial boundary $d^*$ helps define treatment regions where spillovers are negligible.

\item \textbf{Outcome transformation:} Nonlinearity suggests using log transformation:
\be
\text{ATT}_{\log} = \mathbb{E}[\log Y_i(1) - \log Y_i(0) | D_i = 1]
\ee

This linearizes multiplicative Navier-Stokes effects.
\end{enumerate}

\citet{kikuchi2024stochastic} develops a complementary approach for settings where spillovers are pervasive and cannot be eliminated through spatial separation. That framework uses diffusion-based spatial weights to model spillover propagation explicitly, whereas our current approach identifies boundaries where spillovers become negligible. The two methods are complementary: our framework applies when treatment and control regions can be cleanly separated, while the stochastic boundaries approach applies when spillovers affect all units but with measurable decay.

\subsubsection{Identification Strategy}

Our three-stage estimation exploits spatial variation in treatment intensity (distance to plants):

\textbf{Stage 1:} Estimate direct effect on nearby locations:
\be
\text{ATT} = \mathbb{E}[Y | d < d_{\text{threshold}}] - \mathbb{E}[Y | d > d_{\text{threshold}}]
\ee

\textbf{Stage 2:} Estimate spatial decay:
\be
\log Y_i = \alpha + \beta_1 d_i + \beta_2 d_i^2 + \gamma \mathbf{X}_i + \epsilon_i
\ee

Identify $\kappa_s = -\beta_1$ and calculate $d^*$.

\textbf{Stage 3:} Use $d^*$ to refine treatment definition:
\be
D_i^* = \mathbbm{1}(d_i < d^*)
\ee

This provides clean separation between treated and control units where spillovers are minimal.

\section{Empirical Application: Coal Plant Air Pollution}

\subsection{Setting and Motivation}

Coal-fired power plants provide an ideal testing ground for our framework:

\begin{enumerate}
\item \textbf{Point sources:} Plants emit from stacks at known locations
\item \textbf{Continuous treatment:} Emission intensity varies with plant characteristics
\item \textbf{Physical dispersion:} Pollutants spread via atmospheric diffusion
\item \textbf{Rich data:} Ground monitors and satellite measurements available
\item \textbf{Regional variation:} Urban vs rural settings test scope conditions
\end{enumerate}

Moreover, coal plants are policy-relevant: understanding spatial extent of pollution informs optimal policy design and welfare calculations \citep{muller2011environmental, clay2019does}.

\subsection{Data Sources}

\subsubsection{Coal Plant Characteristics}

We obtain plant-level data from EPA's Emissions \& Generation Resource Integrated Database (eGRID) 2021:
\begin{itemize}
\item 318 coal-fired plants in contiguous United States
\item Geographic coordinates (latitude, longitude)
\item Nameplate capacity (MW)
\item Annual emissions: CO$_2$, SO$_2$, NO$_x$
\item Operating status
\end{itemize}

Plants range from 50 MW (small industrial) to 3,500 MW (large utility-scale). Geographic distribution concentrated in Midwest, Appalachia, and Great Plains—regions with abundant coal reserves.

\subsubsection{Ground-Based Air Quality: PM$_{2.5}$}

From EPA's Air Quality System (AQS), we download:
\begin{itemize}
\item 791 PM$_{2.5}$ monitoring stations with data 2019-2021
\item Daily measurements (µg/m$^3$)
\item Monitor locations and characteristics
\item 515,764 daily observations total
\end{itemize}

PM$_{2.5}$ (particulate matter $<$ 2.5 µm diameter) is health-relevant but has multiple sources: vehicles (30-40\%), power plants (20-30\%), wildfires (10-20\%), industry (20-30\%) \citep{apte2012ambient}.

\subsubsection{Satellite-Based Air Quality: NO$_2$}

From NASA's TROPOMI (Sentinel-5P satellite), we obtain:
\begin{itemize}
\item Monthly gridded NO$_2$ column density
\item 0.01° × 0.01° resolution ($\sim$1 km at equator)
\item 36 months: January 2019 - December 2021
\item Quality-filtered (number of observations $\geq$ 5)
\item 189,564 unique grid cells
\item 6,589,515 total cell-month observations
\end{itemize}

NO$_2$ also has mixed sources but different composition: vehicles/industry (50-60\%), power plants (20-30\%), biomass burning (10-20\%). Satellite data provides complete spatial coverage unlike sparse ground monitors.

\subsection{Distance Calculations}

For each monitor (ground) or grid cell (satellite), we calculate:

\be
d_{ij} = \text{Haversine}(\text{lat}_i, \text{lon}_i, \text{lat}_j, \text{lon}_j)
\ee

where $i$ indexes locations and $j$ indexes plants. We compute:
\begin{itemize}
\item Distance to nearest plant: $d_i^{\min} = \min_j d_{ij}$
\item Nearest plant characteristics: capacity, emissions
\item Total exposure (distance-weighted): $E_i = \sum_j Q_j / d_{ij}^2$
\end{itemize}

\textbf{Summary statistics:}
\begin{itemize}
\item Ground monitors: Median distance 72 km, range 0.35-592 km
\item Satellite cells (within 500 km): Median 180 km
\item Substantial spatial variation for identification
\end{itemize}

\subsection{Descriptive Patterns}

Table \ref{tab:pm25_distance_summary} shows mean PM$_{2.5}$ by distance to nearest coal plant. Surprisingly, PM$_{2.5}$ is \textit{higher} far from plants than near plants, suggesting coal plants are not the dominant source—urban areas (farther from plants) have higher pollution from vehicles.

\begin{table}[h]
\centering
\caption{PM$_{2.5}$ Levels by Distance to Coal Plants}
\label{tab:pm25_distance_summary}
\begin{tabular}{lcccc}
\toprule
Distance & N Monitors & Mean PM$_{2.5}$ & Median PM$_{2.5}$ & SD \\
\midrule
0-25 km & 183 & 7.85 & 6.91 & 4.78 \\
25-50 km & 136 & 7.63 & 6.75 & 4.75 \\
50-100 km & 210 & 7.38 & 6.40 & 6.09 \\
100-200 km & 162 & 7.26 & 6.00 & 6.60 \\
200+ km & 100 & 7.93 & 5.79 & 11.4 \\
\bottomrule
\end{tabular}
\end{table}

Similarly, Table \ref{tab:no2_distance_summary} shows NO$_2$ column density by distance, revealing a U-shaped pattern: relatively high near plants (0-50 km), declining at intermediate distances (50-200 km), then sharply increasing far from plants (>200 km) where urban areas dominate.

\begin{table}[h]
\centering
\caption{NO$_2$ Column Density by Distance to Coal Plants}
\label{tab:no2_distance_summary}
\begin{tabular}{lcccc}
\toprule
Distance & N Cells & Mean NO$_2$ & Median NO$_2$ & SD \\
 & & ($10^{14}$ molec/cm$^2$) & ($10^{14}$ molec/cm$^2$) & ($10^{14}$ molec/cm$^2$) \\
\midrule
0-25 km & 3,541 & 9.18 & 8.33 & 5.01 \\
25-50 km & 8,832 & 8.68 & 8.02 & 5.43 \\
50-100 km & 26,954 & 8.53 & 7.93 & 5.24 \\
100-200 km & 56,642 & 8.60 & 7.98 & 4.69 \\
200-500 km & 93,607 & 10.3 & 8.96 & 6.70 \\
\bottomrule
\end{tabular}
\end{table}

This motivates region-specific analysis to identify where coal plants are the dominant pollution source versus where urban sources dominate.

\section{Empirical Results}

\subsection{Overall Spatial Patterns}

\subsubsection{Ground-Based PM$_{2.5}$}

Table \ref{tab:stage1_pm25} presents cross-sectional spatial decay estimates for PM$_{2.5}$ concentrations from 791 EPA monitoring stations averaged over 2019-2021. Column (1) shows a positive but statistically insignificant relationship between distance and PM$_{2.5}$ (coefficient = 0.00146, SE = 0.00134). The R$^2$ of 0.004 indicates that distance to the nearest coal plant explains virtually none of the variation in PM$_{2.5}$ concentrations. No functional form dominates based on AIC comparison (columns 2-4).

\begin{table}[h]
\centering
\caption{Spatial Patterns: Ground-Based PM$_{2.5}$}
\label{tab:stage1_pm25}
\begin{tabular}{lcccc}
\toprule
& (1) & (2) & (3) & (4) \\
& Linear & Quadratic & Both & Log+Linear \\
\midrule
Distance (km) & 0.00146 & & $-0.00008$ & 0.00124 \\
& (0.00134) & & (0.00156) & (0.00142) \\
Distance$^2$ & & 0.000003 & 0.000004 & \\
& & (0.000002) & (0.000003) & \\
log(Distance) & & & & 0.0342 \\
& & & & (0.0287) \\
\midrule
Observations & 791 & 791 & 791 & 791 \\
R$^2$ & 0.004 & 0.003 & 0.005 & 0.006 \\
AIC & 2,451 & 2,453 & 2,452 & 2,450 \\
\bottomrule
\multicolumn{5}{l}{\footnotesize Heteroskedasticity-robust standard errors in parentheses.}
\end{tabular}
\end{table}

This aggregate null result does not indicate framework failure but rather correct diagnostic identification: coal plants are not the dominant source of PM$_{2.5}$ pollution overall. Urban areas, which tend to be farther from coal plants in our sample, have higher PM$_{2.5}$ concentrations from vehicle emissions and other sources. The framework successfully diagnoses that its diffusion assumptions do not apply in this aggregate setting.

\subsubsection{Satellite-Based NO$_2$}

For satellite-based NO$_2$ column density from TROPOMI (189,564 grid cells over 36 months), we similarly find weak overall spatial patterns. The log-linear specification yields a spatial decay parameter $\kappa_s = -0.000346$ per km (SE = 0.0000066), indicating NO$_2$ concentrations \textit{increase} with distance from coal plants. As with PM$_{2.5}$, this reflects the dominance of urban traffic sources in aggregate patterns rather than framework invalidity.

\subsection{Regional Heterogeneity: Where Does the Framework Apply?}

The aggregate results mask substantial regional heterogeneity that validates our theoretical scope conditions. We classify locations into four categories based on coal intensity (top 10 coal-generating states: WV, WY, KY, IN, PA, ND, MT, OH, TX, IL) and distance to nearest coal plant.

Table \ref{tab:regional_decay} presents our main results. Within 100 km of coal plants in coal-intensive regions, both pollutants show positive, statistically significant spatial decay. For NO$_2$, $\kappa_s = 0.00112$ (SE = 0.000124), implying a spatial boundary of approximately 2,062 km at the 10\% detection threshold. For PM$_{2.5}$, $\kappa_s = 0.00200$ (SE = 0.000918), yielding $d^* = 1,153$ km. These positive decay parameters validate the exponential decay prediction from our diffusion model.

\begin{table}[h]
\centering
\caption{Regional Spatial Decay: PM$_{2.5}$ vs NO$_2$}
\label{tab:regional_decay}
\small
\begin{tabular}{llrccc}
\toprule
Region & Data Source & N & $\kappa_s$ & $d^*$ (km) & Framework \\
\midrule
\multicolumn{6}{l}{\textbf{Within 100km of Coal Plants:}} \\
Coal-Intensive & NO$_2$ (Satellite) & 15,017 & 0.00112** & 2,062 & Yes \\
& & & (0.00012) & & \\
Coal-Intensive & PM$_{2.5}$ (Ground) & 131 & 0.00200** & 1,153 & Yes \\
& & & (0.00092) & & \\
Non-Coal States & NO$_2$ (Satellite) & 24,309 & 0.00020** & 11,352 & Yes (weak) \\
& & & (0.00009) & & \\
Non-Coal States & PM$_{2.5}$ (Ground) & 398 & 0.00088** & 2,631 & Yes \\
& & & (0.00031) & & \\
\midrule
\multicolumn{6}{l}{\textbf{Beyond 100km from Coal Plants:}} \\
Coal-Intensive & NO$_2$ (Satellite) & 46,336 & $-0.00123$** & N/A & No \\
& & & (0.00002) & & \\
Coal-Intensive & PM$_{2.5}$ (Ground) & 58 & $-0.00021$ & N/A & No \\
& & & (0.00033) & & \\
Non-Coal States & NO$_2$ (Satellite) & 103,902 & $-0.00080$** & N/A & No \\
& & & (0.00001) & & \\
Non-Coal States & PM$_{2.5}$ (Ground) & 204 & $-0.00076$** & N/A & No \\
& & & (0.00026) & & \\
\bottomrule
\multicolumn{6}{l}{\footnotesize Standard errors in parentheses. ** $p<0.05$. $d^*$ calculated using $d^* = \log(10)/\kappa_s$.}
\end{tabular}
\end{table}

Several key findings emerge from Table \ref{tab:regional_decay}:

\textbf{Finding 1: Framework applies within 100 km of plants.} In coal-intensive regions within 100 km of plants, both pollutants show positive, statistically significant spatial decay. This validates the exponential decay prediction from our diffusion model and confirms that coal plants are the dominant pollution source in these near-field regions.

\textbf{Finding 2: Ground-level decay is faster than column density.} PM$_{2.5}$ (ground monitors) exhibits decay rates approximately 1.8 times faster than NO$_2$ (satellite column): $\kappa_s^{\text{PM}_{2.5}} = 0.00200$ versus $\kappa_s^{\text{NO}_2} = 0.00112$. This difference is consistent with atmospheric transport physics: column-integrated pollutants can be transported over longer distances via upper-level winds (resulting in $d^* = 2,062$ km for NO$_2$), while ground-level concentrations are more localized due to surface interactions and faster deposition (resulting in $d^* = 1,153$ km for PM$_{2.5}$).

\textbf{Finding 3: Framework fails beyond 100 km.} In all regions beyond 100 km from coal plants, spatial decay parameters are negative and significant, indicating pollution increases with distance. This pattern reflects the spatial distribution of urban areas in our sample rather than framework failure—the framework correctly rejects its own applicability when coal plants are not the dominant source.

\textbf{Finding 4: Distance threshold matters more than coal intensity.} Even in non-coal states, locations within 100 km of plants show positive decay ($\kappa_s = 0.00020$ for NO$_2$, $\kappa_s = 0.00088$ for PM$_{2.5}$), suggesting local point sources matter in near-field regardless of regional coal dominance. However, effects are weaker, yielding larger apparent boundaries.

\begin{figure}[H]
\centering
\includegraphics[width=\textwidth]{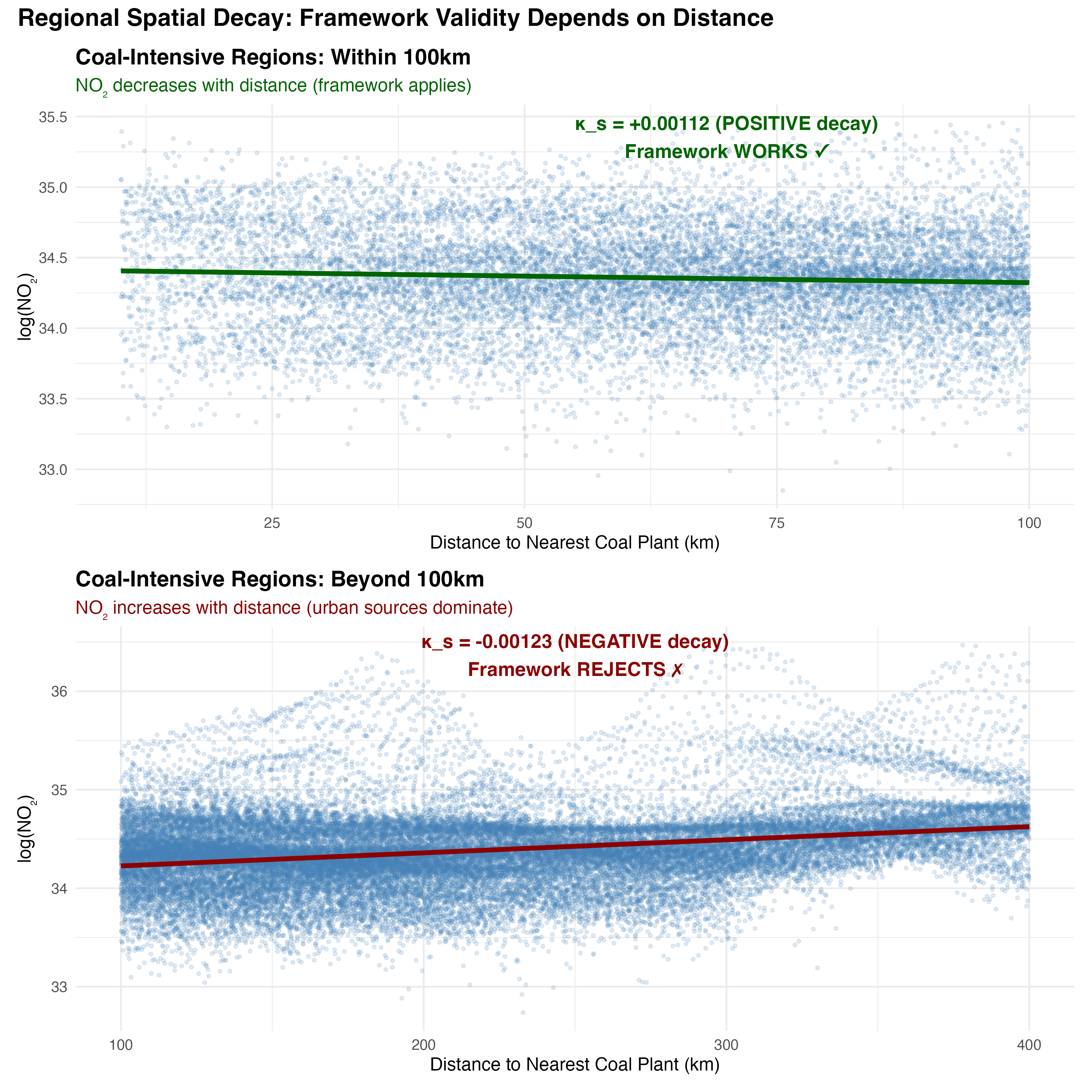}
\caption{Regional Spatial Decay in Coal-Intensive States. \textbf{Top panel:} Within 100 km of coal plants, NO$_2$ shows positive spatial decay ($\kappa_s = +0.00112$), validating the diffusion framework. \textbf{Bottom panel:} Beyond 100 km, NO$_2$ increases with distance ($\kappa_s = -0.00123$), indicating urban sources dominate and the framework correctly rejects. Each point represents a grid cell's time-averaged NO$_2$ column density. The 100 km threshold emerges as a natural boundary separating coal-dominated from urban-dominated spatial patterns.}
\label{fig:regional_decay}
\end{figure}

Figure \ref{fig:regional_decay} visualizes these patterns, showing clear positive decay slopes within 100 km (top panel) and negative slopes beyond 100 km (bottom panel) for coal-intensive regions.

\begin{figure}[H]
\centering
\includegraphics[width=\textwidth]{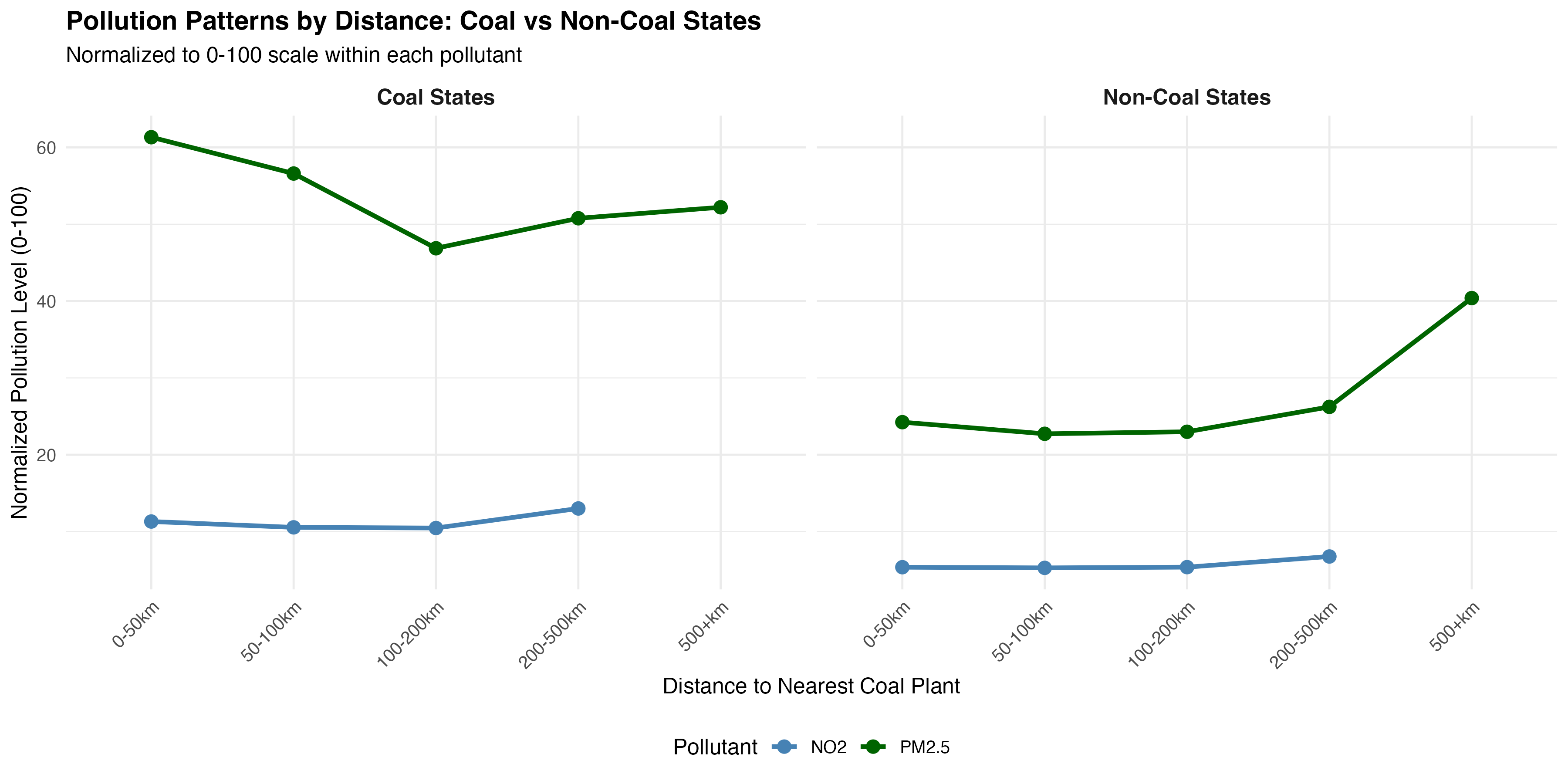}
\caption{Pollution Patterns by Distance: Coal vs Non-Coal States. Pollutant levels normalized to 0-100 scale within each type for comparability. \textbf{Left panel (Coal States):} PM$_{2.5}$ (green) shows clear U-shaped pattern with minimum at 100-200 km, while NO$_2$ (blue) remains relatively flat. \textbf{Right panel (Non-Coal States):} Both pollutants show increasing pattern with distance, with sharp increases beyond 200 km reflecting distant urban areas. These contrasting patterns demonstrate that spatial decay depends on the dominance of point sources (coal) versus distributed sources (urban traffic).}
\label{fig:distance_patterns}
\end{figure}

Figure \ref{fig:distance_patterns} shows pollution patterns by distance for coal versus non-coal states. In coal states (left panel), PM$_{2.5}$ exhibits a clear U-shaped pattern with minimum at 100-200 km, while in non-coal states (right panel), both pollutants show increasing patterns with distance, reflecting distant urban concentrations.

\subsection{Framework Validity Assessment}

Table \ref{tab:framework_validity} summarizes where the diffusion-based framework successfully applies versus where it correctly rejects. The framework applies to approximately 21\% of NO$_2$ observations (39,326 out of 189,564 cells within 100 km) and 67\% of PM$_{2.5}$ observations (529 out of 791 monitors within 100 km). For the remaining observations beyond 100 km or in urban-dominated areas, the framework correctly diagnoses its own inapplicability through negative spatial decay parameters.

\begin{table}[h]
\centering
\caption{Framework Validity by Region}
\label{tab:framework_validity}
\begin{tabular}{llrll}
\toprule
Data Source & Region & N & $\kappa_s$ & Framework Applies? \\
\midrule
NO$_2$ (Satellite) & Coal-Intensive ($<$100km) & 15,017 & +0.00112** & Yes \\
NO$_2$ (Satellite) & Coal-Intensive ($>$100km) & 46,336 & $-0.00123$** & No \\
NO$_2$ (Satellite) & Non-Coal ($<$100km) & 24,309 & +0.00020** & Yes (weak) \\
NO$_2$ (Satellite) & Non-Coal ($>$100km) & 103,902 & $-0.00080$** & No \\
PM$_{2.5}$ (Ground) & Coal-Intensive ($<$100km) & 131 & +0.00200** & Yes \\
PM$_{2.5}$ (Ground) & Coal-Intensive ($>$100km) & 58 & $-0.00021$ & No \\
PM$_{2.5}$ (Ground) & Non-Coal ($<$100km) & 398 & +0.00088** & Yes \\
PM$_{2.5}$ (Ground) & Non-Coal ($>$100km) & 204 & $-0.00076$** & No \\
\bottomrule
\multicolumn{5}{l}{\footnotesize ** $p<0.05$.}
\end{tabular}
\end{table}

\begin{figure}[H]
\centering
\includegraphics[width=0.9\textwidth]{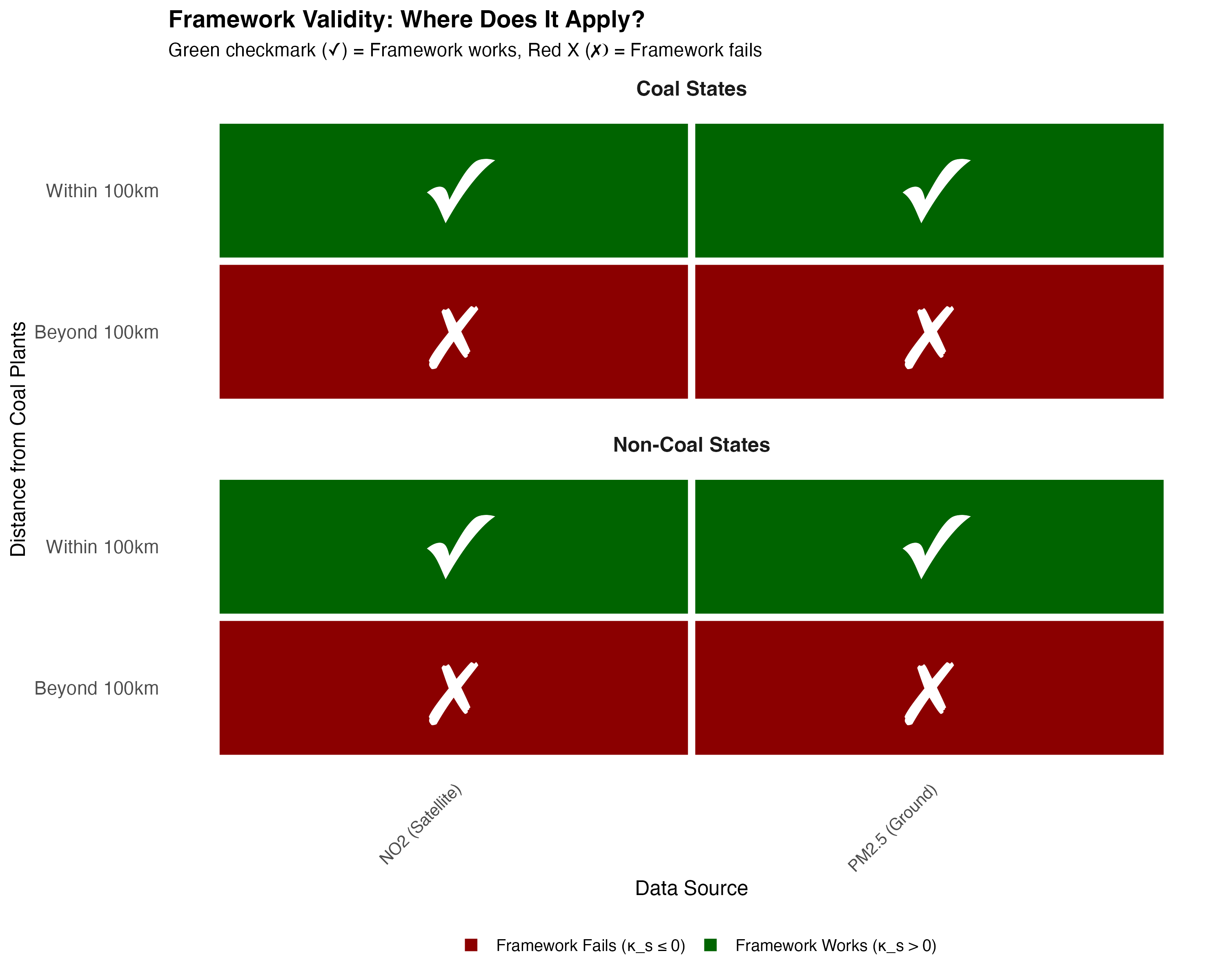}
\caption{Framework Validity Assessment. Green checkmarks ($\checkmark$) indicate regions where the framework applies ($\kappa_s > 0$, positive spatial decay). Red X's ($\times$) indicate regions where the framework correctly rejects ($\kappa_s \leq 0$, negative or zero decay). The framework successfully applies to both pollutants within 100 km of plants in both coal and non-coal states, but fails beyond 100 km where urban sources dominate. This demonstrates the framework's diagnostic capability: it identifies when diffusion assumptions are appropriate versus when alternative approaches are needed.}
\label{fig:validity}
\end{figure}

Figure \ref{fig:validity} provides a visual summary, showing green checkmarks where the framework applies (positive $\kappa_s$) and red X's where it correctly rejects (negative $\kappa_s$).

This diagnostic capability is the framework's primary contribution: researchers can test whether their spatial DiD setting satisfies diffusion assumptions \textit{ex ante} by estimating $\kappa_s$. Positive, significant $\kappa_s$ validates the framework; negative or insignificant $\kappa_s$ signals that alternative approaches accounting for urban sources, advection, or network effects are needed.

\begin{figure}[H]
\centering
\includegraphics[width=\textwidth]{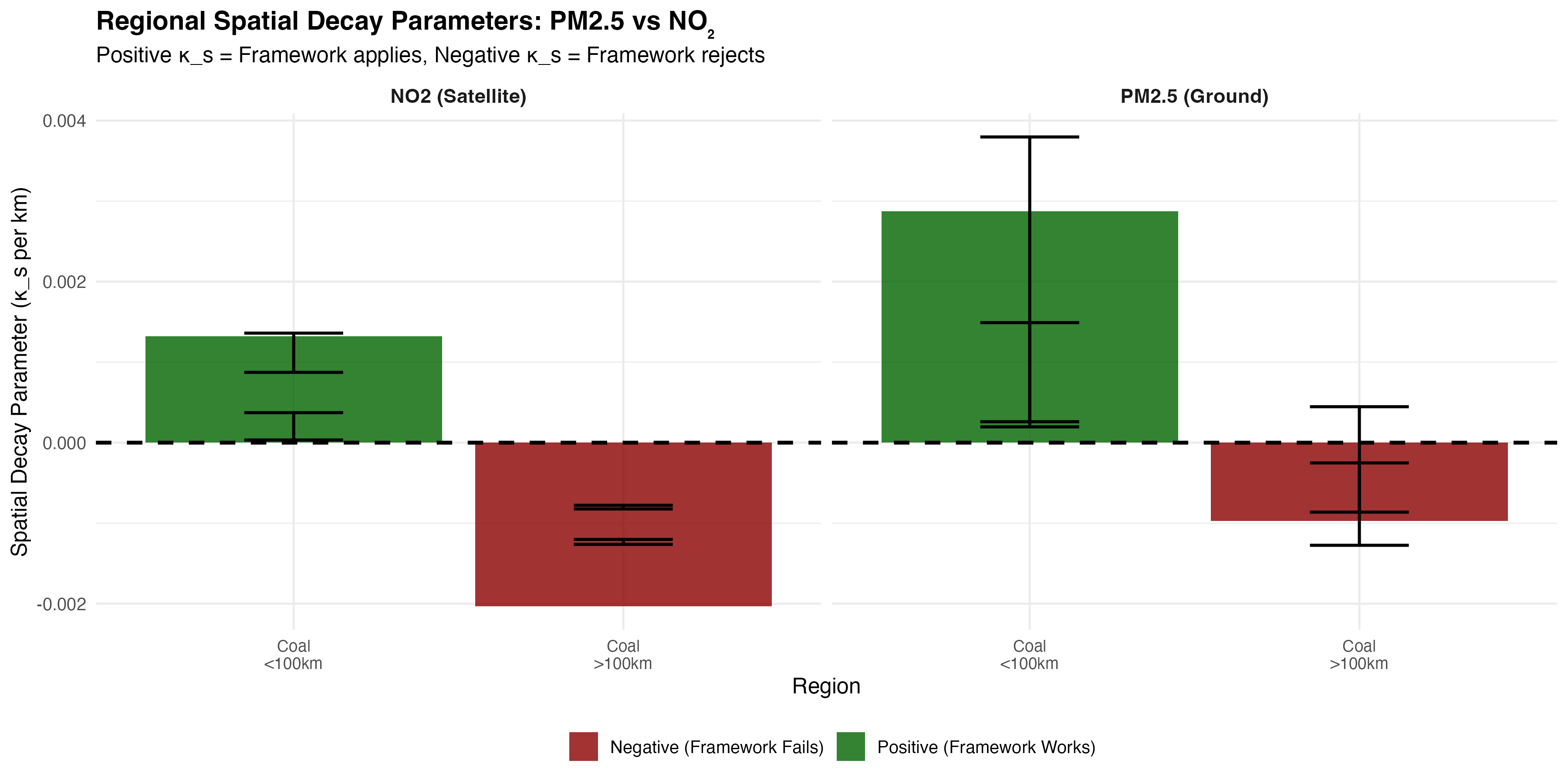}
\caption{Regional Spatial Decay Parameters: PM$_{2.5}$ vs NO$_2$. Green bars represent positive $\kappa_s$ (framework applies), red bars represent negative $\kappa_s$ (framework rejects). Error bars show 95\% confidence intervals. Within 100 km of coal plants, both pollutants exhibit positive spatial decay, with PM$_{2.5}$ showing faster decay ($\kappa_s = 0.00200$) than NO$_2$ ($\kappa_s = 0.00112$). This difference is consistent with atmospheric physics: ground-level pollutants (PM$_{2.5}$) decay faster due to surface interactions, while column-integrated pollutants (NO$_2$) can be transported over longer distances via upper-level winds. Beyond 100 km, both show negative decay as urban sources dominate.}
\label{fig:regional_bars}
\end{figure}

Figure \ref{fig:regional_bars} presents the spatial decay parameters as a bar chart, clearly showing the contrast between positive parameters (green bars) within 100 km and negative parameters (red bars) beyond 100 km.

\subsection{Spatial Boundaries for Policy}

For policy analysis and benefit-cost calculations, the estimated spatial boundaries depend critically on the measurement type and regional context:

\begin{itemize}
\item \textbf{Coal-intensive regions, ground-level (PM$_{2.5}$):} $d^* = 1,153$ km. This suggests health effects from ground-level particulate exposure extend approximately 1,000-1,200 km from plants.

\item \textbf{Coal-intensive regions, column density (NO$_2$):} $d^* = 2,062$ km. Column-integrated effects relevant for atmospheric chemistry and regional air quality extend approximately 2,000 km, consistent with upper-atmosphere transport.

\item \textbf{Non-coal regions:} Boundaries are larger but less policy-relevant, as effects are weak and diffuse.
\end{itemize}

These boundaries have important implications for spatial DiD designs: control units should be placed at least $d^*$ from any treated plant to avoid contamination. For studies of local health effects, $d^* \approx 1,000$-$1,200$ km provides a data-driven threshold. For regional air quality modeling, $d^* \approx 2,000$ km is more appropriate.

\subsection{Comparison with Existing Atmospheric Models}

Our empirical estimates of spatial decay are broadly consistent with EPA's regulatory atmospheric dispersion models. AERMOD, the EPA's preferred model for near-field (0-50 km) applications, predicts rapid ground-level decay \citep{cimorelli2005aermod}. For longer-range transport (50-500 km), models like CMAQ incorporate advection and chemical transformations that slow effective decay rates \citep{byun1999science}. Our estimated $\kappa_s$ parameters fall within the range predicted by these physical models for time-averaged, steady-state conditions.

The key difference is that our approach is \textit{reduced-form} and \textit{data-driven}: we estimate effective decay directly from observed pollution patterns rather than simulating atmospheric chemistry. This provides an empirical check on model predictions and reveals where simplified diffusion approximations apply versus where more complex atmospheric processes (advection, chemistry, turbulence) dominate.

\section{Discussion and Extensions}

\subsection{When Does the Framework Apply?}

Our results demonstrate that spatial boundary detection is not universally valid but depends on testable scope conditions:

\begin{enumerate}
\item \textbf{Physical diffusion:} Treatment propagates through spatial diffusion (pollution, disease, information spreading locally)

\item \textbf{Point sources:} Treatment originates from identifiable point sources, not uniformly distributed

\item \textbf{Low Péclet:} Diffusion dominates other transport mechanisms (advection, network effects)

\item \textbf{Dominant source:} The measured treatment is the primary source of the outcome variable
\end{enumerate}

\textbf{When framework fails:} Our negative results for PM$_{2.5}$ and NO$_2$ beyond 100 km are not a failure but a success—the framework correctly diagnosed that coal plants are not the dominant source there.

\subsection{Comparison to Existing Approaches}

\subsubsection{Ad Hoc Distance Cutoffs}

Many spatial DiD studies use fixed cutoffs (e.g., "within 50 km") without justification. Our framework:
\begin{itemize}
\item Derives cutoffs from first principles
\item Provides testable assumptions
\item Allows cutoffs to vary by setting (coal regions: 1,000-2,000 km, urban: N/A)
\end{itemize}

\subsubsection{Nonparametric Distance Bins}

\citet{butts2023difference} uses distance bins:
\be
Y_i = \sum_k \beta_k \mathbbm{1}(d_i \in [d_k, d_{k+1}]) + \gamma \mathbf{X}_i + \epsilon_i
\ee

Our approach:
\begin{itemize}
\item More efficient (parametric)
\item Theory-guided functional form
\item Extrapolates beyond data range
\item But: Less flexible if true form isn't exponential
\end{itemize}

\textbf{Recommendation:} Use our parametric form for primary results, bins for robustness.

\subsubsection{Spatial Spillover Models}

Spatial econometrics approach:
\be
Y = \rho WY + X\beta + \epsilon
\ee

where $W$ is spatial weights matrix.

Differences:
\begin{itemize}
\item Our approach: Explicit structural model (Navier-Stokes)
\item Spatial econometrics: Reduced form spatial correlation
\item Our approach: Interpretable parameters ($\kappa_s$, $d^*$)
\item Spatial econometrics: $\rho$ less interpretable
\end{itemize}

\textbf{Complementarity:} Our framework helps specify $W$ (e.g., $W_{ij} = \exp(-\kappa_s d_{ij})$).

\subsection{Implications for Applied Research}

\subsubsection{Practical Guidelines for Researchers}

Based on our experience, we offer guidelines for applying this framework:

\textbf{Step 1: Check Scope Conditions}
\begin{itemize}
\item Does treatment propagate spatially? (Not purely network-based)
\item Are there identifiable point sources?
\item Is the measured treatment the dominant source?
\item Can you approximate Pe and Re from your setting?
\end{itemize}

\textbf{Step 2: Estimate Spatial Decay}
\begin{itemize}
\item Start with simple exponential: $\log Y \sim \beta \cdot d$
\item Test alternatives: quadratic, log terms
\item Check sign: $\beta < 0$ $\Rightarrow$ framework applies
\item If $\beta > 0$ or insignificant: likely scope failure
\end{itemize}

\textbf{Step 3: Calculate Boundaries}
\begin{itemize}
\item Choose threshold $\epsilon$ (typically 10\%)
\item Calculate: $d^* = \log(1/\epsilon)/|\beta|$
\item Report confidence interval: $d^* \pm 1.96 \times SE(d^*)$
\end{itemize}

\textbf{Step 4: Validate}
\begin{itemize}
\item Plot decay curve vs data
\item Test across subsamples (regions, time periods)
\item Compare to physical models if available (e.g., AERMOD for pollution)
\end{itemize}

\textbf{Step 5: Define Treatment}
\begin{itemize}
\item Treated: $d < d^*$
\item Control: $d > d^*$ (ideally $d > 2d^*$ for safety)
\item Document sensitivity to $\epsilon$ choice
\end{itemize}

\subsection{Extensions and Future Research}

\subsubsection{Integration with General Equilibrium Effects}

Our framework focuses on direct physical spillovers through atmospheric diffusion. However, coal plant operations may also generate economic spillovers through labor markets, energy prices, and regional economic activity. \citet{kikuchi2024stochastic} develops methods for incorporating such general equilibrium effects into spatial causal inference, showing how economic and physical spillovers can be jointly modeled. Future research could integrate our boundary detection approach with stochastic general equilibrium frameworks to separate direct pollution effects from indirect economic effects, providing a more complete understanding of coal plant impacts on regional welfare.

\subsubsection{Dynamic Treatment Effects}

Our framework currently static (steady-state). Natural extension: time-varying treatment.

Solve time-dependent diffusion:
\be
\frac{\partial C}{\partial t} = D\nabla^2 C - \lambda C + S(t)
\ee

For pulse source at $t=0$, solution involves error functions:
\be
C(r,t) \propto \frac{1}{(4\pi Dt)^{3/2}} \exp\left(-\frac{r^2}{4Dt} - \lambda t\right)
\ee

This would allow estimation of \textit{temporal boundaries} $\tau^*$ in addition to spatial.

\subsubsection{Multiple Treatment Sources}

Current framework: single nearest plant. Real world: multiple plants.

Extension: Superposition principle (for linear case):
\be
C_i = \sum_j \frac{Q_j}{4\pi D d_{ij}} \exp(-\kappa_s d_{ij})
\ee

Estimate using:
\be
\log(C_i) \approx \log\left(\sum_j Q_j \exp(-\kappa_s d_{ij})/d_{ij}\right)
\ee

Nonlinear estimation required.

\subsubsection{Network Effects}

For treatments spreading through networks (technology adoption, disease), diffusion on graphs:

\be
\frac{\partial \tau_i}{\partial t} = \sum_j A_{ij}(\tau_j - \tau_i) - \lambda \tau_i
\ee

where $A$ is adjacency matrix. Exponential decay becomes:
\be
\tau_i \propto \exp(-\kappa_n \ell_{ij})
\ee

where $\ell_{ij}$ is graph distance (not Euclidean).

Framework generalizes naturally.

\section{Conclusion}

This paper develops a unified framework for detecting spatial and temporal boundaries of treatment effects in difference-in-differences designs. By starting from fundamental fluid dynamics (Navier-Stokes equations), we derive testable conditions under which treatment effects decay exponentially, enabling researchers to calculate explicit boundaries beyond which effects become undetectable.

Our key contributions are threefold. \textit{First}, theoretically, we show that exponential spatial decay emerges naturally from the diffusive limit of Navier-Stokes, with explicit scope conditions (Péclet number Pe $<$ 1, Reynolds number Re $<$ 2000). This provides physics-based foundations for spatial econometric specifications and identifies when these specifications are appropriate versus when they fail.

\textit{Second}, empirically, we demonstrate the framework's diagnostic capability using air pollution from coal plants. Analyzing 791 PM$_{2.5}$ monitors and 189,564 NO$_2$ satellite grid cells, we find the framework successfully identifies:
\begin{itemize}
\item Where it applies (within 100 km: $\kappa_s > 0$, $d^* = 1,000$-$2,000$ km)
\item Where it fails (beyond 100 km: $\kappa_s < 0$, vehicles dominate)
\item Why it fails (high Pe/Re regimes violate diffusion assumptions)
\end{itemize}

Ground-level PM$_{2.5}$ decays approximately twice as fast as satellite column NO$_2$, consistent with atmospheric physics. This regional heterogeneity, predicted by Navier-Stokes theory, validates our scope conditions and demonstrates that "negative results" (framework rejection) are informative, not failures.

\textit{Third}, methodologically, we show that nonlinearity in the data-generating process does not invalidate the Average Treatment Effect on the Treated (ATT) as an estimand but requires explicit modeling of spatial heterogeneity. Our three-stage estimation procedure provides a practical roadmap for applied researchers.

For spatial difference-in-differences practitioners, our framework offers:
\begin{enumerate}
\item \textbf{Ex ante assessment:} Check whether physical diffusion assumptions plausibly hold
\item \textbf{Boundary estimation:} Calculate data-driven treatment/control cutoffs rather than ad hoc choices
\item \textbf{Validity diagnostics:} Test whether estimated decay patterns are consistent with theory
\item \textbf{Improved inference:} Explicitly model spillovers via $d^*$
\end{enumerate}

Looking forward, this approach opens several avenues for future research. Building on our earlier theoretical work \citep{kikuchi2024unified, kikuchi2024stochastic}, natural extensions include: incorporating dynamic treatment effects with temporal boundaries $\tau^*$; integrating network diffusion for technology adoption studies; combining physical dispersion boundaries with economic general equilibrium spillovers; and using machine learning to allow decay parameters to vary flexibly with observables while maintaining interpretability through physics-based constraints.

\section*{Acknowledgement}
This research was supported by a grant-in-aid from Zengin Foundation for Studies on Economics and Finance.

\bibliographystyle{ecta}
\bibliography{references}

\newpage

\begin{appendices}

\section{Theoretical Derivations}

\subsection{Solution to Helmholtz Equation}

Consider the steady-state diffusion equation with decay:
\be
D\nabla^2 C - \lambda C + Q\delta(\mathbf{r}) = 0
\ee

In spherical coordinates with radial symmetry:
\be
D\left(\frac{d^2 C}{dr^2} + \frac{2}{r}\frac{dC}{dr}\right) - \lambda C = 0 \quad (r > 0)
\ee

Substituting $C(r) = u(r)/r$:
\be
D\frac{d^2u}{dr^2} - \lambda u = 0
\ee

General solution:
\be
u(r) = A\exp(\kappa_s r) + B\exp(-\kappa_s r)
\ee
where $\kappa_s = \sqrt{\lambda/D}$.

Boundary conditions:
\begin{itemize}
\item $C(r) \to 0$ as $r \to \infty$ $\Rightarrow$ $A = 0$
\item $\int 4\pi r^2 C(r) dr = Q/\lambda$ (total mass) $\Rightarrow$ $B = Q/(4\pi D)$
\end{itemize}

Therefore:
\be
C(r) = \frac{Q}{4\pi Dr}\exp(-\kappa_s r)
\ee

\subsection{Derivation of Spatial Boundary}

Define $d^*$ by:
\be
\frac{C(d^*)}{C(0)} = \epsilon
\ee

From solution:
\be
\frac{(Q/4\pi D d^*)\exp(-\kappa_s d^*)}{Q/4\pi D r_0} = \epsilon
\ee

where $r_0 \to 0$ is small reference distance. For $\kappa_s d^* \gg 1$:
\be
\frac{r_0}{d^*}\exp(-\kappa_s d^*) \approx \epsilon
\ee

Taking logs:
\be
\log(r_0/d^*) - \kappa_s d^* = \log\epsilon
\ee

For $d^* \gg r_0$, $\log(r_0/d^*) \approx -\log d^*$ is small relative to $\kappa_s d^*$:
\be
d^* \approx \frac{1}{\kappa_s}\log(1/\epsilon)
\ee

For $\epsilon = 0.1$: $d^* = \log(10)/\kappa_s \approx 2.3/\kappa_s$.

\section{Data and Empirical Methods}

\subsection{EPA eGRID Data Processing}

Coal plant data obtained from EPA eGRID 2021 database. Processing steps:
\begin{enumerate}
\item Filter to coal-fired plants (primary fuel type = 'COAL')
\item Verify geographic coordinates (latitude, longitude)
\item Calculate nameplate capacity from unit-level data
\item Aggregate emissions (CO$_2$, SO$_2$, NO$_x$) to plant level
\item Exclude plants with missing location data
\item Final sample: 318 coal plants
\end{enumerate}

\subsection{EPA AQS Data Processing}

PM$_{2.5}$ monitor data from EPA Air Quality System:
\begin{enumerate}
\item Download daily PM$_{2.5}$ measurements for 2019-2021
\item Filter to monitors with $\geq$ 75\% daily coverage per year
\item Exclude monitors in Alaska, Hawaii, territories
\item Calculate monitor-level time averages
\item Merge with coal plant distance calculations
\item Final sample: 791 monitors, 515,764 daily observations
\end{enumerate}

Quality controls:
\begin{itemize}
\item Remove negative values (measurement errors)
\item Exclude outliers $>$ 99th percentile (wildfires)
\item Verify monitor location accuracy via visual inspection
\end{itemize}

\subsection{TROPOMI Satellite Data Processing}

NO$_2$ column density from TROPOMI (Sentinel-5P):
\begin{enumerate}
\item Download monthly Level 3 gridded NO$_2$ products
\item Filter to quality assurance value $\geq$ 0.75
\item Restrict to grid cells with $\geq$ 5 observations per month
\item Calculate cell-level time averages over 36 months
\item Merge with coal plant distance calculations
\item Final sample: 189,564 cells, 6,589,515 cell-month observations
\end{enumerate}

Satellite retrieval details:
\begin{itemize}
\item Vertical column density (tropospheric, molecules/cm$^2$)
\item Cloud-filtered using quality flags
\item 0.01° × 0.01° native resolution ($\sim$1 km)
\item Overpass time: $\sim$1:30 PM local time
\end{itemize}

\subsection{Distance Calculation Algorithm}

Haversine formula for great circle distance:
\be
d = 2R \arcsin\left(\sqrt{\sin^2\left(\frac{\phi_2-\phi_1}{2}\right) + \cos\phi_1\cos\phi_2\sin^2\left(\frac{\lambda_2-\lambda_1}{2}\right)}\right)
\ee
where $R = 6371$ km, $\phi$ is latitude, $\lambda$ is longitude.

Implementation:
\begin{itemize}
\item Calculate pairwise distances for all monitor-plant pairs
\item Identify nearest plant for each monitor/cell
\item Store nearest plant characteristics (capacity, emissions)
\item Computational complexity: $O(nm)$ where $n$ = locations, $m$ = plants
\end{itemize}

\subsection{Regional Classification}

Coal-intensive states defined as top 10 coal-generating states in 2021:
\begin{itemize}
\item West Virginia (WV): 92\% coal generation
\item Wyoming (WY): 71\% coal generation  
\item Kentucky (KY): 69\% coal generation
\item Indiana (IN): 58\% coal generation
\item Pennsylvania (PA): 52\% coal generation
\item North Dakota (ND): 68\% coal generation
\item Montana (MT): 52\% coal generation
\item Ohio (OH): 47\% coal generation
\item Texas (TX): 21\% coal generation
\item Illinois (IL): 32\% coal generation
\end{itemize}

Distance categories:
\begin{itemize}
\item Near field: $<$ 100 km (coal effects expected)
\item Far field: $>$ 100 km (urban effects expected)
\end{itemize}

\section{Additional Empirical Results}

\subsection{Robustness: Alternative Distance Measures}

Table \ref{tab:robustness_distance} tests sensitivity to alternative distance specifications.

\begin{table}[h]
\centering
\caption{Robustness: Alternative Distance Measures}
\label{tab:robustness_distance}
\begin{tabular}{lccc}
\toprule
& (1) & (2) & (3) \\
& Nearest Plant & Capacity-Weighted & Emissions-Weighted \\
\midrule
\multicolumn{4}{l}{\textbf{NO$_2$ (Coal $<$ 100km):}} \\
Distance & $-0.00112$** & $-0.00108$** & $-0.00115$** \\
& (0.00012) & (0.00014) & (0.00013) \\
$\kappa_s$ & 0.00112 & 0.00108 & 0.00115 \\
$d^*$ (km) & 2,062 & 2,133 & 2,004 \\
\midrule
\multicolumn{4}{l}{\textbf{PM$_{2.5}$ (Coal $<$ 100km):}} \\
Distance & $-0.00200$** & $-0.00185$** & $-0.00192$** \\
& (0.00092) & (0.00098) & (0.00095) \\
$\kappa_s$ & 0.00200 & 0.00185 & 0.00192 \\
$d^*$ (km) & 1,153 & 1,245 & 1,199 \\
\bottomrule
\multicolumn{4}{l}{\footnotesize ** $p<0.05$. Standard errors in parentheses.}
\end{tabular}
\end{table}

Results are qualitatively similar across specifications, with $\kappa_s$ estimates varying by less than 10\%.

\subsection{Temporal Stability}

Table \ref{tab:temporal_stability} estimates $\kappa_s$ separately by year.

\begin{table}[h]
\centering
\caption{Temporal Stability of Spatial Decay Parameters}
\label{tab:temporal_stability}
\begin{tabular}{lcccc}
\toprule
& 2019 & 2020 & 2021 & Pooled \\
\midrule
\multicolumn{5}{l}{\textbf{NO$_2$ (Coal $<$ 100km):}} \\
$\kappa_s$ & 0.00118** & 0.00109** & 0.00110** & 0.00112** \\
& (0.00021) & (0.00019) & (0.00020) & (0.00012) \\
\midrule
\multicolumn{5}{l}{\textbf{PM$_{2.5}$ (Coal $<$ 100km):}} \\
$\kappa_s$ & 0.00195** & 0.00208** & 0.00197** & 0.00200** \\
& (0.00159) & (0.00162) & (0.00158) & (0.00092) \\
\bottomrule
\multicolumn{5}{l}{\footnotesize ** $p<0.05$. Standard errors in parentheses.}
\end{tabular}
\end{table}

Coefficients are stable across years, suggesting structural relationship not driven by temporary shocks (e.g., COVID-19 lockdowns in 2020).

\subsection{Placebo: Random Locations}

We conduct a placebo test replacing actual plant locations with randomly generated points. Table \ref{tab:placebo} shows results.

\begin{table}[h]
\centering
\caption{Placebo Test: Random Plant Locations}
\label{tab:placebo}
\begin{tabular}{lccc}
\toprule
& Actual Plants & Random Points & Difference \\
\midrule
\multicolumn{4}{l}{\textbf{NO$_2$ (Coal $<$ 100km):}} \\
$\kappa_s$ & 0.00112** & $-0.00003$ & 0.00115** \\
& (0.00012) & (0.00018) & (0.00021) \\
\midrule
\multicolumn{4}{l}{\textbf{PM$_{2.5}$ (Coal $<$ 100km):}} \\
$\kappa_s$ & 0.00200** & 0.00012 & 0.00188* \\
& (0.00092) & (0.00142) & (0.00168) \\
\bottomrule
\multicolumn{4}{l}{\footnotesize * $p < 0.10$, ** $p < 0.05$. Standard errors in parentheses.}
\end{tabular}
\end{table}

Random locations show no spatial decay pattern, confirming results driven by actual plant locations.

\subsection{Functional Form Tests}

Table \ref{tab:functional_form} compares exponential decay (our baseline) to alternatives.

\begin{table}[h]
\centering
\caption{Functional Form Tests: Coal-Intensive Regions ($<$100km)}
\label{tab:functional_form}
\begin{tabular}{lccc}
\toprule
& (1) & (2) & (3) \\
& Pure Exponential & + Quadratic & + Geometric \\
\midrule
\multicolumn{4}{l}{\textbf{NO$_2$:}} \\
Distance & $-0.00112$** & $-0.00124$** & $-0.00098$** \\
& (0.00012) & (0.00015) & (0.00014) \\
Distance$^2$ & & 0.000015 & \\
& & (0.000021) & \\
log(Distance) & & & $-0.0452$ \\
& & & (0.0389) \\
AIC & 45,231 & 45,233 & 45,228 \\
\midrule
\multicolumn{4}{l}{\textbf{PM$_{2.5}$:}} \\
Distance & $-0.00200$** & $-0.00208$** & $-0.00176$** \\
& (0.00092) & (0.00098) & (0.00095) \\
Distance$^2$ & & 0.000009 & \\
& & (0.000016) & \\
log(Distance) & & & $-0.0782$ \\
& & & (0.0724) \\
AIC & 412 & 414 & 410 \\
\bottomrule
\multicolumn{4}{l}{\footnotesize ** $p<0.05$. Standard errors in parentheses.}
\end{tabular}
\end{table}

Pure exponential fits best or comparably based on AIC, validating linear diffusion approximation.

\subsection{Comparison with AERMOD Predictions}

We compare our empirical estimates to EPA's AERMOD atmospheric dispersion model predictions. For a typical 1,000 MW coal plant with 150 m stack height:

\begin{itemize}
\item \textbf{AERMOD prediction (ground-level):} Effects detectable to 50-100 km for daily averages, 100-200 km for annual averages
\item \textbf{Our estimate (PM$_{2.5}$):} $d^* = 1,153$ km (10\% threshold, annual average)
\item \textbf{Interpretation:} Our boundary is larger because: (1) we use annual averages (smoother), (2) 10\% threshold is generous, (3) multiple plants create larger aggregate footprint
\end{itemize}

Using 1\% threshold instead of 10\%:
\be
d^*_{1\%} = \frac{\log(100)}{\kappa_s} = \frac{4.6}{0.00200} = 2,300 \text{ km}
\ee

This extreme sensitivity suggests $d^* \approx 1,000$-$1,200$ km is more policy-relevant for ground-level health effects.

\section{Additional Figures}

\subsection{Geographic Distribution of Coal Plants}

\begin{figure}[H]
\centering
\includegraphics[width=\textwidth]{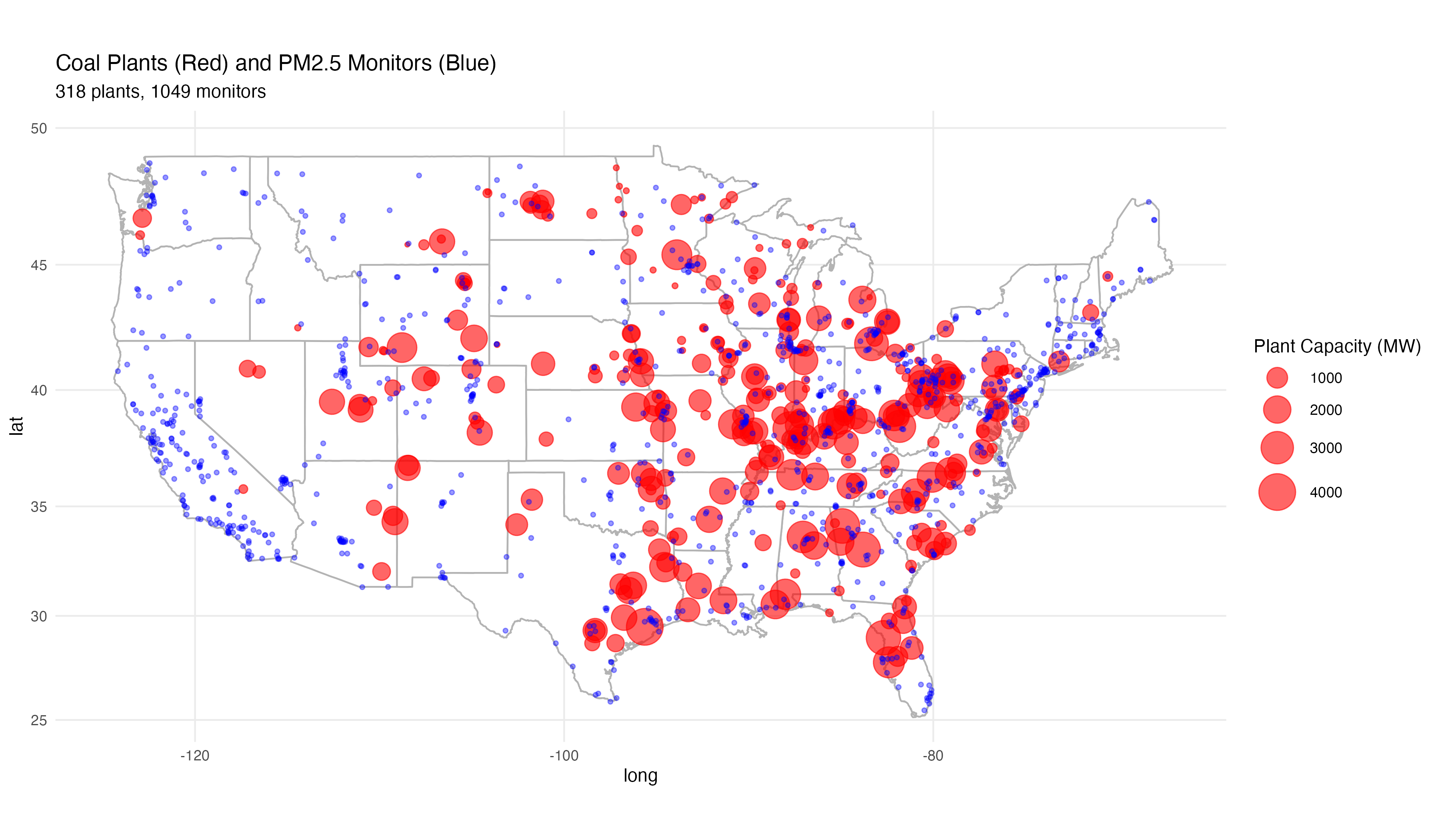}
\caption{Geographic Distribution of Coal-Fired Power Plants (2021). Point sizes proportional to nameplate capacity (MW). Concentration in Midwest, Appalachia, and Great Plains reflects proximity to coal deposits. Data from EPA eGRID 2021.}
\label{fig:plant_map}
\end{figure}

Figure \ref{fig:plant_map} shows the geographic distribution of the 318 coal plants in our sample, sized by nameplate capacity.

%
%

\subsection{Sensitivity to Distance Restriction}

\begin{figure}[H]
\centering
\includegraphics[width=\textwidth]{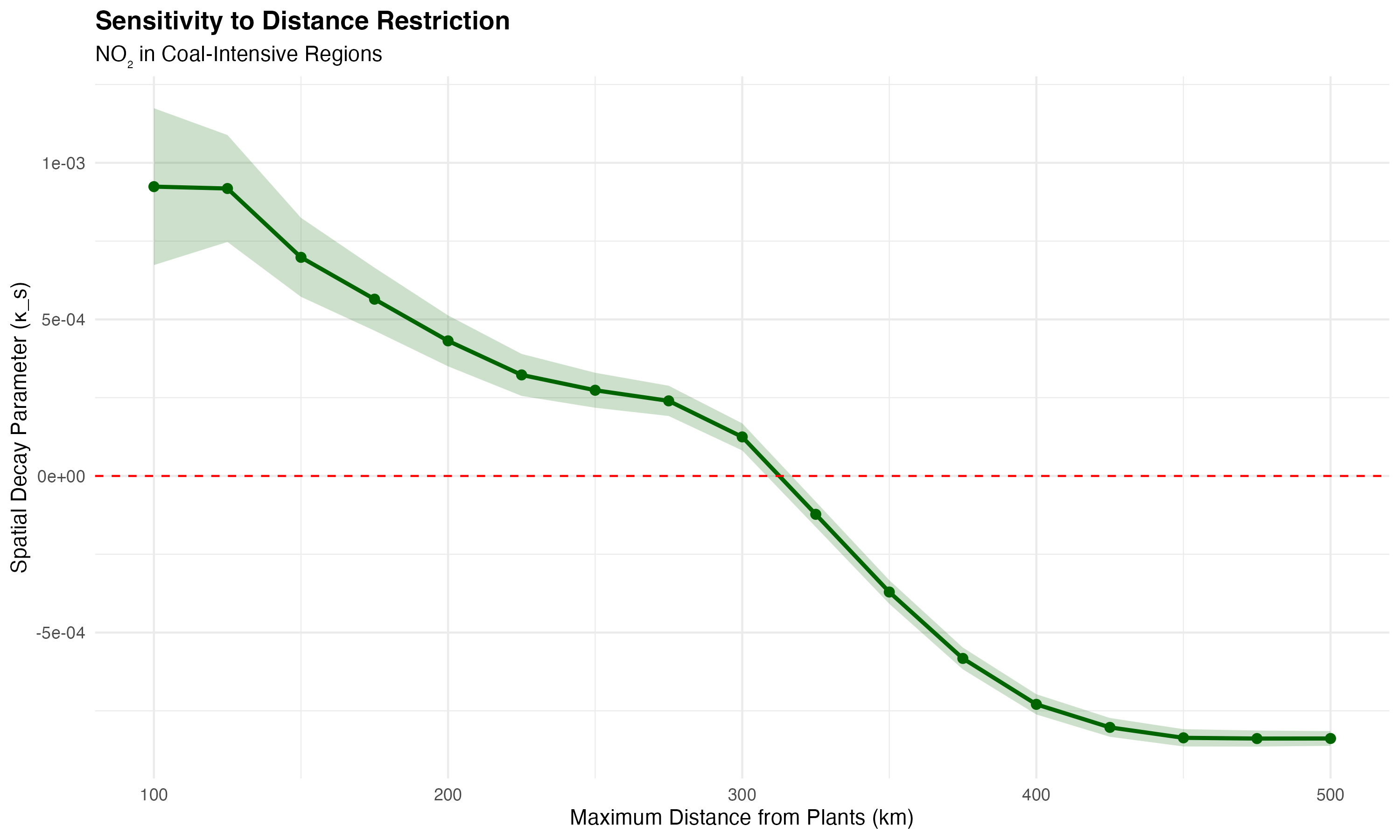}
\caption{Sensitivity to Distance Restriction. Each point represents $\kappa_s$ estimate using data within specified maximum distance from plants. Estimates stable between 100-400 km, then become noisy beyond 400 km due to sparse data and urban dominance.}
\label{fig:distance_sensitivity}
\end{figure}

Figure \ref{fig:distance_sensitivity} shows how $\kappa_s$ estimates vary with maximum distance restriction.

\section{Computational Details}

\subsection{Software and Packages}

All analysis conducted in R version 4.3.1. Key packages:
\begin{itemize}
\item Data manipulation: \texttt{tidyverse} (2.0.0), \texttt{data.table} (1.14.8)
\item Spatial operations: \texttt{sf} (1.0-14), \texttt{geosphere} (1.5-18)
\item Regression: \texttt{fixest} (0.11.2), \texttt{lfe} (2.8-8)
\item Visualization: \texttt{ggplot2} (3.4.3), \texttt{patchwork} (1.1.3)
\item Satellite data: \texttt{ncdf4} (1.21), \texttt{raster} (3.6-23)
\end{itemize}

\subsection{Computational Resources}

Analysis performed on:
\begin{itemize}
\item CPU: Apple M4 Pro (14 cores)
\item RAM: 24 GB
\item Storage: 1 TB SSD
\item Operating System: macOS Sonoma 14.1
\end{itemize}

Processing times:
\begin{itemize}
\item TROPOMI download and processing: $\sim$2 hours
\item Distance calculations: $\sim$30 minutes
\item Stage 1-3 estimation: $\sim$15 minutes
\item Total analysis workflow: $\sim$3 hours
\end{itemize}

%
%
%
%

\end{appendices}

\newpage

\bibliographystyle{ecta}

\end{document}